# Holographic gratings for cold neutron optics

M. Fally[1], J. Klepp[1], C. Pruner[2], Y. Tomita[3], H. Eckerlebe[4], J. Kohlbrecher[5], R.A. Rupp[1]

[1]University of Vienna, Faculty of Physics, A-1090 Wien, Austria;martin.fally@univie.ac.at
[2]University of Salzburg, Department of Materials Science and Physics, A-5020 Salzburg, Austria
[3]University of Electro-Communications, Department of Electronics Engineering,1-5-1 Chofugaoka, Chofu, Tokyo 182, Japan
[4]Helmholtz-Zentrum Geesthacht, D-21502 Geesthacht, Germany
[5]Paul Scherrer Institut, CH-5232 Villigen PSI, Switzerland

We discuss the applicability of holographically patterned polymers, polymer dispersed liquid crystals, and nanoparticle-polymer composites as optical elements for cold neutrons. Requirements concerning the spacing, thickness or strength of the grating for certain types of neutron optical elements, e.g., 2-port or 3-port beamsplitters, are discussed in the framework of a rigorous coupled-wave analysis. Finally, a roadmap to neutron mirrors, e.g., for interferometers, is drawn.

1. Introduction

The successful demonstration of neutron-optical phenomena such as reflection, diffraction or interference has led to devices, which are able to manipulate neutron beams [1]. Such instruments monochromate, collimate or polarize neutrons for further use in fundamental research as well as applications in materials science, biology or crystallography. Employing an interferometer it is even possible to measure relative phases between two states probing the influence of fundamental interaction potentials [2]. The efficiency of such neutron-optical instruments is determined by the neutron-matter interaction described by the neutron-optical potential $V$ or equivalently the neutron refractive-index $n = \sqrt{1-V/E}$ at an incident-neutron energy $E$. Introducing the coherent scattering length density $b\rho$ – a characteristic quantity for a certain material – the refractive index can be shown to depend quadratically on the wavelength $\lambda$

$$n_o = 1 - \lambda^2\, b\rho/(2\pi). \qquad (1)$$

Thus it is obvious, that larger wavelength neutrons yield stronger effects. In contrast to thermal neutrons, where perfect single crystals are used for interferometry [3], gratings with a proper lattice constant $\Lambda$ were lacking for cold neutrons.

During the last two decades we have tried to tackle this problem by using a holographic technique in photosensitive polymers and polymer composites to fabricate efficient diffraction gratings for cold neutrons. We demonstrated that by combining such gratings an interferometer for cold neutrons can be successfully operated [4].

In this contribution we discuss various possibilities for fabricating diffraction gratings with high efficiency.

2. Holographically Prepared Gratings For Neutrons and Diffraction Regimes

To prepare holographic gratings we expose a photosensitive medium to the interference pattern of two plane waves originating from coherent laser light. The resulting spatial sinusoidal intensity variation $I(x)$ is mapped to the refractive index (for neutrons) $n(x)=n_o+\Delta n\cos(x\, 2\pi/\Lambda)$ with $\Lambda$ in the order of the light wavelength (several hundred nanometers). The light-induced refractive-index modulation $\Delta n = \lambda^2\, b\Delta\rho/(2\pi)$ is one of the decisive parameters towards highly efficient gratings. Keeping in mind the applicability for a Mach-Zehnder type interferometer [4], gratings should operate in the two-beam coupling (=Bragg) regime and reach reflectivities of 50% for beam splitters and 100% for mirrors. There are two experimental constraints in reaching these goals: the limits given by the light optical preparation of the grating such as the thickness $d$, $\Lambda$ and $b\Delta\rho$ as well as the restrictions implied by the neutron optic setup (the wavelength $\lambda$ and its distribution $\Delta\lambda$, the divergence $\Delta\theta$ of the neutron beam). For an optimal grating the angular width of the reflectivity curve should be considerably larger than the divergence of the beam, i.e., $\Lambda/d > \Delta\theta$, or in other words: for a given spacing $\Lambda$ the thickness should be chosen rather thin. On the other hand, to reach the desired reflectivities, the grating modulation parameter $\nu = d\lambda b\Delta\rho/2$ should be around 1, a value that can most easily be reached by making the gratings rather thick. To decide about an optimal combination of the accessible parameters ($d$, $\Lambda$, $b\Delta\rho$) we make use of a rigorous coupled wave analysis (RCWA) [5] and discuss the dependence of the first order reflectivity for cold neutrons on the Klein-Cooke parameter $Q=2\pi d\lambda/\Lambda^2$ and the grating modulation parameter $\nu$.

Three different types of photosensitive materials were tested for their usability in neutron optics: deuterated polymethylmethacrylate (d-PMMA), holographic polymer dispersed liquid crystals (HPDLC) and polymer nanoparticle composites (PnpC). The figure of merit for the gratings recorded in these media is given by their light-induced refractive-index modulation for neutrons, i.e. basically by the coherent scattering length density modulation $b\Delta\rho$. While for the pure polymer only number density modulation occurs, the latter two can also be tuned via the species of the liquid crystalline compound or the type of nanoparticles ($SiO_2$ and $ZrO_2$ in our case). Basic parameters are summarized in Table 1, details can be found in [4,6-8].

|  | $d$ (μm) | $\Lambda$ (μm) | $b\Delta\rho$ (1/μm²) |
|---|---|---|---|
| d-PMMA | (1...3)×10³ | 0.25...0.8 | N/A |
| HPDLC | 10...30 | 0.4...1.2 | 2...10 |
| PnpC $SiO_2$ | 50...100 | 0.5; 1 | 1...6 |
| PnpC $ZrO_2$ | 50 | 1 | 3 |

Table 1: Accessible parameter ranges for diffraction gratings in various materials

### 3. Neutron Experiments & Results

Neutron experiments were performed on various small angle neutron scattering (SANS) beamlines in Germany, Switzerland and France. Typical collimation of about 20 m with slits of a few millimeters ensured a divergence of about a mrad. The diffracted signal was detected by a two-dimensional detector at a distance of again 20 m from the sample. Typical cold wavelengths between 1 and 2 nm with a spread of 10% were used. To measure the reflectivity curves, the samples were placed on a rotation stage. Fig. 1 shows the results for d-PMMA, HPDLC and PnpC $SiO_2$ as examples.

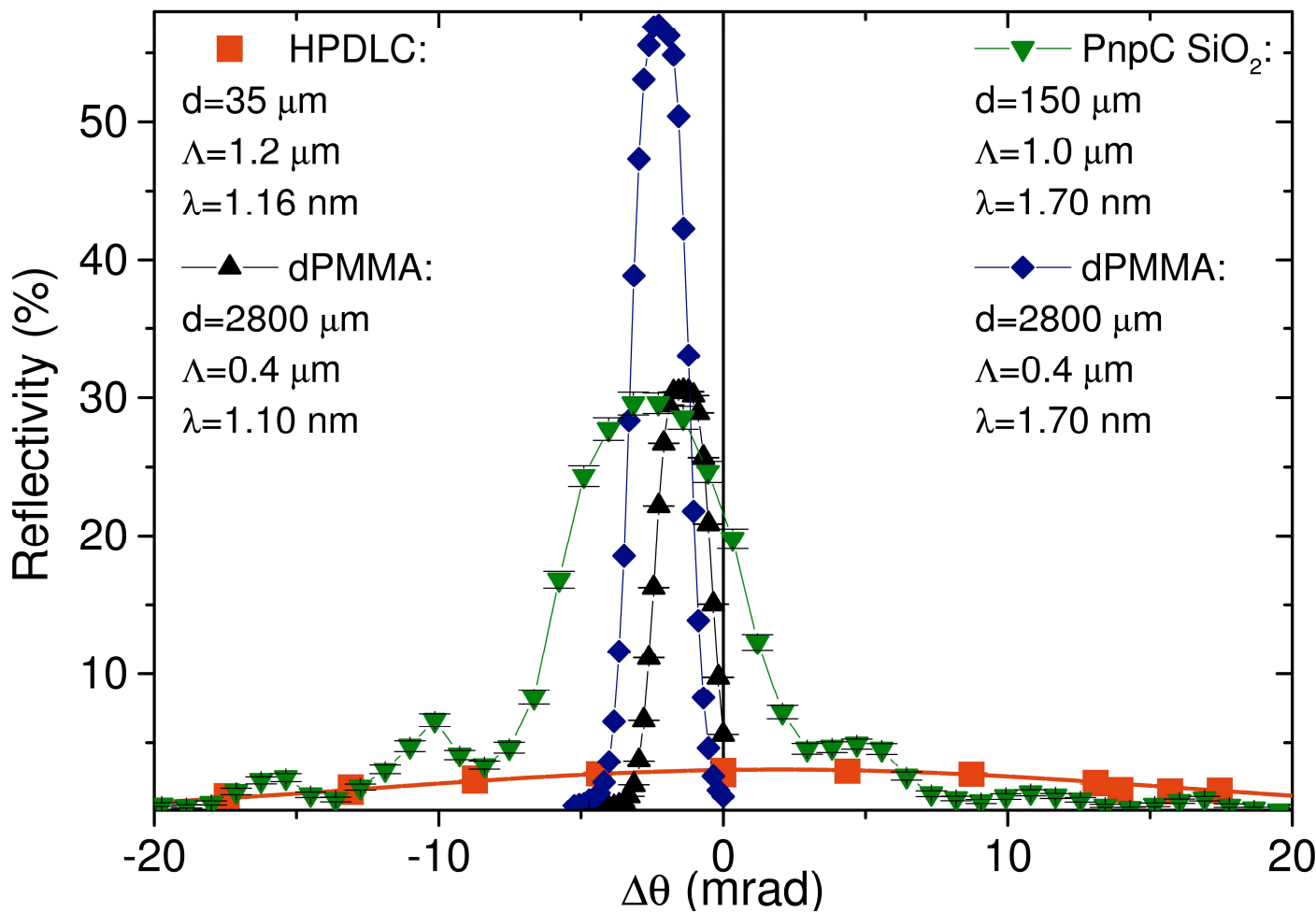

Figure 1: Reflectivity (=rocking curves) for gratings with largely varying parameters $d$, $\Lambda$ and $\lambda$ as given in the legend.

## 4. Discussion

The search for proper conditions to realize beamsplitters and mirrors for cold neutrons forces us to fabricate a grating with the following properties: geometrically as thin as possible, high reflectivity and exhibiting diffraction in the Bragg regime. These conditions can be visualized using the RCWA and depicting the reflectivity as a function of $Q$ and $\nu$ (see Fig. 2).

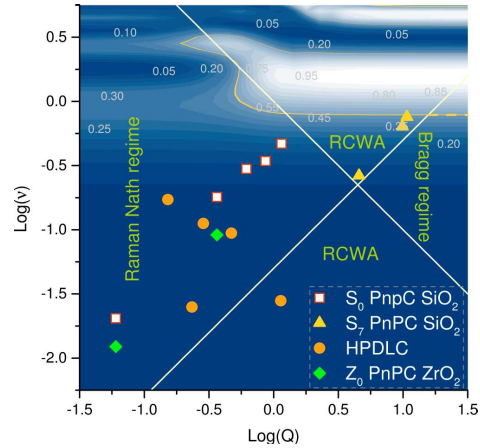

Figure 2: Contour plot for the first-order reflectivity in the $\nu$-$Q$ parameter space. Yellow lines indicate 50% reflectivity. Experimental data points for various materials and conditions are included.

In this double-logarithmic contour plot an increase of $b\Delta\rho$ means vertical motion up, a decrease of $\Lambda$ horizontal motion to the right, and changing thickness or wavelength a diagonal motion (see definition of $\nu$ and $Q$ above). Unfortunately, from an experimental point of view these parameters cannot be tuned independently. It is well known that a decrease of the grating spacing also reduces the refractive-index modulation [9] and that an increase of the (grating-) thickness, for some materials, is not possible due to strongly enhanced scattering [10]. However, by recording gratings in PnpC at a moderate thickness (~100 μm) and tilting them by an angle $\zeta$ around the grating vector, so that the effective thickness increases to $d/\cos(\zeta)$, a beamsplitter could be put into practice [8].

## 5. Conclusions and Outlook

We gave an outline on how to produce diffraction gratings for cold neutron optical purposes with high reflectivities by employing a holographic technique.

Very recently we started studying superparamagnetic nanoparticles embedded in polymers. Upon application of a magnetic field the nuclear scattering term as indicated by the coherent scattering length in (1) is altered by a contribution of magnetic scattering. Then we will be able to suppress scattering for one spin state and find enhanced diffraction for the other. Such diffraction gratings are switchable by means of magnetic fields and could serve as polarizing beam splitters and/or be used as analyzers instead of the unhandy and expensive $He^3$-filters.

## 6. Acknowledgements

This work is based on experiments performed at the Swiss spallation neutron source SINQ, Paul Scherrer Institute, Villigen, Switzerland. Financial support by the Austrian Science Fund (FWF): P18988 and P20265 as well as the ÖAD-WTZ SI 07/2011. **Dedicated to W. Ellmeyer on the occasion of his 50th birthday.**